\documentclass[journal]{IEEEtran}
\usepackage{amsfonts}
\usepackage{graphicx}
\usepackage{amsmath}
\usepackage[caption=false]{subfig}
\ifCLASSINFOpdf
\else
\fi
\begin{document}
%
\title{Principled Data Completion of Network Constraints for Day Ahead Auctions in Power Markets}
%
%
%

\author{Ioan Alexandru Puiu,~\IEEEmembership{University of Oxford,}
        Raphael Andreas Hauser,~\IEEEmembership{University of Oxford}
\thanks{I.A Puiu and R.A. Hauser are with the Mathematical Institute, University of Oxford, UK, e-mail: ioan.puiu@maths.ox.ac.uk and (see https://www.maths.ox.ac.uk/people/raphael.hauser)}
\thanks{Funding: I.A.Puiu was supported by the EPSRC CDT in Industrially Focused Mathematical Modelling in collaboration with Macquarie Group.}
\thanks{This work has been submitted to the IEEE for possible publication. Copyright may be transferred without notice, after which this version may no longer be accessible.}}

%
%

\markboth{}%
{Shell \MakeLowercase{\textit{et al.}}: Bare Demo of IEEEtran.cls for IEEE Journals}
%



\maketitle

\begin{abstract}
Network constraints play a key role in the price finding mechanism for European Power Markets, but historical data is very sparse and usually insufficient for many quantitative applications. We reconstruct the constraints data, known as the Power Transmission Distribution Factors (PTDFs) and Remaining Available Margins (RAMs), by first recovering the underlying time dependent signals known as the Generation Shift Keys (GSKs) and Phase Angles (PAs), and the electricity grid characteristics, via a mathematical optimisation problem. This is solved by exploiting marginal convexity in certain subspaces via alternating minimisation. The GSKs and PAs are then mapped to the PTDFs and RAMs, using the grid structure. Our reconstruction achieves in-sample  relative errors of $22.3\%$ and $15.9\%$ for the PTDFs and RAMs respectively, while the out-of-sample ones are $25.4\%$ and $18.1\%$ respectively. We further show that our model outperforms the naive approach, and that the reconstructed GSKs and PAs recover specific structure.
\end{abstract}


%
\IEEEpeerreviewmaketitle

\section{Introduction and Background Information}
\IEEEPARstart{P}{ower} markets are frequently organised as Day Ahead Auctions (DAA) due to the difficulty of
storing electricity, coupled with uncertainty in demand and production patterns. We consider the Flow Based Market Coupling (FBMC) approach, which uses physical principles for tracking the flow loads in the grid. This has become increasingly popular since it generally offers increased operational safety, trading volume and social welfare, when compared to the Available Transfer Capacity (ATC) approach \cite{ACER}. FBMC was pioneered in the Central Western European (CWE) Power Market, which we use as a case study for our methodology.
\par
The core of the FBMC approach is a set of linear constraints enforced at each time step on the zonal (country) production levels, which ensure that the electricity grid always operates within the safety limits. This set of constraints governs the feasibility domain of power production and flow patterns, and thus ultimately determines zonal prices and their potential decoupling. However, only a very small subset of these constraints is published \textit{ex-post} \cite{JAO}, which is generally insufficient for domain reconstruction or thorough statistical analysis that accounts for the data complexity. Further, these constraints change rather erratically in time making naive techniques inappropriate for sound quantitative decision making. For the DAA participants, this translates to reduced ability to manage risk, and ultimately to increased market price inefficiencies.
\par
To address this, we propose a methodology that uses the physical principles of the FBMC approach and publicly available data to reconstruct the full set of linear constraints, known as the combination of Power Transmission Distribution Factors (PTDFs) and Remaining Available Margins (RAMs). To this end, we first recover the electricity grid structure, as well as the underlying time-dependent signals known as the Generation Shift Keys (GSKs) and Phase Angles (PAs). The GSKs and PAs are then transformed to the PTDFs and RAMs using maps determined by the graph structure. We achieve the reconstruction by formulating an optimisation model based on the principles of FBMC, regularised by the available prior knowledge. We exploit the structure of the optimisation problem by solving via alternating minimisation: stochastic optimisation is used for the non-linear and computationally intensive parts of the objective, and convex quadratic programming for the marginally convex components. We are able to show good results for the reconstruction of PTDFs and RAMs in out-of-sample tests, outperforming the naive approach. Further, we analyse the structure of the recovered grid characteristics, as well as of the GSKs and PAs signals, and we show that we recover specific clustering structure, {suggesting good reconstruction quality}.
\par
The novel contribution of our approach is fourfold: (i) a graph model containing the quantitative physical characteristics is obtained, which is useful for any fundamental model that relies on the graph structure, (ii) the underlying GSKs and PAs signals are recovered, which give insight into the transmission system operators' behaviour, (iii) full reconstruction of the PTDFs and RAMs, allowing for full feasibility domain analysis by market participants, and (iv) our approach enables simulation and forecasting based on the reconstructed signals. 
\par
The rest of the paper is structured as follows: background information is presented in Section II and a brief discussion of the Market Data is given in Section III. Sections IV \& V present the main optimisation model and its algorithmic solution. Numerical results are presented in Section VI, and conclusions are drawn in Section VII.
\section{Background Information}
Each set of power injections and extractions decisions determines the flows across the lines of the network, via Kirchhoff’s Law. Transmission system operators (TSO) monitor grid transmission lines (and other elements), and ensure that power flows do not exceed the safety limits by imposing the linear constraints given by the $(PTDF,RAM)$ combination \cite{JAODocumentation}. The constraints calculations in FBMC are based on the DC flow model presented in detail in \cite{WPEN2014}, which involves a linear approximation based on three assumptions: (i) there are no losses on transmission lines, i.e. the resistance on lines is zero, (ii) the voltage profile is flat, i.e. all nodes have the same voltage, and (iii) the voltage angle differences are small for neighbouring nodes. 
\par
Let the power grid be represented by a graph $\mathcal{G}$, with a transmission lines set $\mathcal{L}$ of cardinality $|\mathcal{L}| = L$, and node set $\mathcal{N}$ with $|\mathcal{N}|=N$ nodes. Let the zone set for the considered graph be $\mathcal{Z}$ with $|\mathcal{Z}|=Z$, and the node set of each zone $z$ be $\mathcal{N}_z$ with $|\mathcal{N}_z| = N_z$. The nodal PTDF is fully determined by the physical characteristics of the power grid and is given by
\begin{equation}\label{nodal PTDF}
	P(b,A)=B\tilde{A} (\tilde{A}^{T}B\tilde{A})^{-1}, 
\end{equation}
where $\tilde{A}=A\Omega_{ref}^T$, with $A \in \mathbb{R}^{L \times N}$ being the line-node incidence matrix of the power grid defined as 
\begin{equation}\label{Adef}
A_{ji}=
	\left\{\begin{array}{ll}
	+1&\text{  if line $j$ originates in node $i$}\\
	-1&\text{  if line $j$ terminates at node $i$}\\
	\textrm{ }0&\textrm{ }\textrm{     otherwise}
	\end{array} \right., 
\end{equation}
and $\Omega_{ref} \in \mathbb{R}^{(N-1)\times N}$ is a projection matrix obtained by deleting row $i_0$ of the identity matrix $I_N \in \mathbb{R}^{N \times N}$, with $i_0$ some chosen reference node \cite{WPEN2014}. Further, the diagonal susceptance matrix $B = \textrm{diag}(b)$ aggregates the line susceptabces $b_j$, $\forall j \in \mathcal{L}$. The nodal power changes $\Delta p_N = p_N - g^0 \in \mathbb{R}^{N}$ are then mapped through the Nodal PTDF to line power flow changes $ \Delta p_L = p_L - F^{ref} \in \mathbb{R}^L$ by
\begin{equation}\label{lineflow}
    \Delta p_L = P(b,A) \Delta p_N,
\end{equation}
where $\Delta p_N$ and $ \Delta p_L$ are defined with respect to $(g^0, F^{ref})$, known as the \textit{base case}, which we describe later. Negative and positive signs in $p_N$ indicate nodal consumption and production respectively. The interpretation of $P_{ji}(b,A)$ is the change in flow along line $j$ that is caused by injection of one unit of power at node $i$ for extraction at the reference node $i_0$. 
\par
Note that the directionality of $A$ and the choice of the reference node $i_0$ are arbitrary but necessary \cite{WPEN2014}, with the only requirement that these are fixed once and for all. Similarly, the phase angles $\alpha^L$ are mapped to power flow changes through the Phase Shift Distribution Factor (PSDF) \cite{WPEN2014}, defined as
\begin{equation}\label{PSDF}
    PSDF(b,A) := B - B\tilde{A}(\tilde{A}^TB\tilde{A})^{-1}\tilde{A}^TB,
\end{equation}
which yields the linear mapping $\Delta^{\alpha} p_L = PSDF(b,A)\alpha^L$,
where $\Delta^{\alpha} p_L$ is the line power changes given by the line phase angles $\alpha^L$. Note that this linearisation is accurate only for sufficiently small $\alpha^L$.
\par
 To enforce the constraints, in each zone, the TSOs define the \textit{base case} which includes a nodal power pattern \cite{JAODocumentation}. We here denote this pattern as $g^0_{z,t} \in \mathbb{R}^{N_z}$. By collating these vectors for all zones $z \in \mathbb{Z}$, one obtains $g^0_t = ||^V_{z \in \mathcal{Z}}\left(g^0_{z,t}\right) \in \mathbb{R}^N$, where $||^V_.(.)$ represents vertical concatenation. The \textit{base case} is defined $ex-ante$ and is the TSO's best guess of the nodal production and consumption patterns sitting at day $D-2$, for each hour $t$ of day $D$. This is implicitly used in the DAA welfare maximisation problem for price finding and order acceptance at day $D-1$ \cite{Euphemia}. Note that unless otherwise specified, from here onward, $t$ denotes hourly time indexing.
\par
Constraining the DAA auction at nodal level is not practical, as power is not traded on a per node (power-plant) basis. Instead, nodes are collated into zones, and power injection at each node within a zone (usually country) is treated as equivalent and traded at a zonal price. However, it is not possible to reduce the power injection decisions at all nodes within a zone to a single equivalent zonal power injection decision, because flows within each zone must also be kept feasible.
Nevertheless, the impact can be approximated: for each hourly delivery period, the nodal power levels within zone $z$ are assumed to be described by a vector
\begin{equation}\label{powerlinearapprox}
    p^N_{z,t} = g^0_{z,t} +\Delta P_{z,t} g^1_{z,t}, 
\end{equation}
where $\Delta P_{z,t} \in \mathbb{R}$ is the overall power deviation from the \textit{base case} scenario and $g^1_{z,t} \in \mathbb{R}^{N_z}$ is the Generation Shift Key vector representing the power injection at each node in zone $z$ at time $t$, under a shift scenario where one extra unit of power is injected across zone $z$ and extracted at the reference node $i_0$. Similarly to $g^0_{z,t}$, the GSK is also defined at day $D-2$, but obeys $1^Tg^1_{z,t} = 1$, and we define the extended version of $g^1_{z,t}$ for all nodes $\mathcal{N}$ as $\overline{g}^1_{z,t}$ with, $(\overline{g}^1_{z,t})_i=0$, $\forall i \notin \mathcal{N}_z$, instead of restraining $\overline{g}^1_{z,t}$ to the $N_z$ elements in zone $z$. 
\par
Thus, the deviations of the power generation pattern from the base case are modelled as a 1-dimensional set, which is clearly a massive simplification. However, this can be remedied by considering a set of outage scenarios $\mathcal{S}$ for the same hour slot. Most commonly, an
outage $s$ refers to a line $j$, which effectively accounts for removing entry $b_j$ and row $A_{j,:}$, and we denote the resulting quantities as $b^s$ and $A^s$. We restrict our attention only to line outages here. For each $s \in \mathcal{S}$ and $t \in \mathcal{T}$ the \textit{line-zonal} PTDF can be now obtained as \cite{ACERExplanatorynote}
\begin{equation}\label{lineZonalPTDF}
    PTDF_{s,t} = P_s(b,A)G^1_{s,t},
\end{equation}
where the subcript $s$ denotes explicit scenario dependence, $P_s(b,A):= P(b^s, A^s)$, and $G^1_{s,t} = ||^H_{z \in \mathcal{Z}}(\overline{g}^1_{s,t,z}) \in \mathbb{R}^{N \times Z}$ is the horizontal concatenation of the GSKs for each zone under scenario $s$ at time $t$. The constraints can now be written as 
\begin{equation}\label{fmaxconstraints}
    r_{s,t} \leq PTDF_{s,t}\delta p_t^Z \leq R_{s,t},
\end{equation}
where the RAMs are defined as
\begin{align}\label{RAM eqs}
    r_{s,t} = -F^{max}_{s,t} - F^{ref}_{s,t} + FRM_{s,t} - FAV_{s,t} \\
    R_{s,t} =F^{max}_{s,t} - F^{ref}_{s,t} - FRM_{s,t} - FAV_{s,t}
\end{align}
 and the \textit{Net Position} $\delta p_t^Z := ||^V_{z \in \mathcal{Z}}(\Delta P_{z,t}) \in \mathbb{R}^Z$ is the vector of zonal power changes, with respect to the base case values given by $1^T g^0_{z,t}$ \cite{JAO}. By writing $PSDF_s(b,A):=PSDF(b^s,A^s)$, the reference flows are expressed as 
 \begin{equation}\label{Frefeq}
     F_{s,t}^{ref} = P_s(b,A)g^0_{s,t} + {V_0}^2 PSDF_s(b,A)\alpha^L_t,
 \end{equation}
where $V_0$ is the flat voltage profile \cite{WPEN2014}. The FRMs are introduced to account for TSO’s grid modelling errors, which mainly
includes the linearisation simplification but also flow measurements uncertainties \cite{JAO}. They are computed for each line as a multiple of the standard deviation of the error series defined as the difference between measured and computed flows. The final adjustment values (FAV) incorporates TSO information not accounted for in the DC model \cite{JAO}. Finally $F^{max}_{s,t} \in \mathbb{R}^L$ is the vector of maximum allowed flows, which is generally constant in time but can vary when grid conditions change.
\par
Ensuring that every single line $l \in \mathcal{L}$ satisfies the feasibility constraints \eqref{fmaxconstraints} is computationally intensive and somewhat redundant \cite{JAO}. Thus, the TSOs monitor a much smaller set of \textit{critical branches} (CBs), denoted as $\mathcal{L}_{cb} \subset \mathcal{L}$ with cardinality $|\mathcal{L}_{cb}|=L_{cb}$ \cite{ACERExplanatorynote}. Let $\Omega_{cb} \in \mathbb{R}^{L_{cb} \times L}$ be the canonical projection from the space of all lines, $\mathcal{L}$, to the critical ones, $\mathcal{L}_{cb}$. The constraints can then be re-written by multiplying equation \eqref{fmaxconstraints} by $\Omega_{cb}$, and while the form is preserved, the number of constraints reduces dramatically.
\par
The critical branches constraints version of \eqref{fmaxconstraints} are then used in conjunction to the Welfare Maximisation Problem (WMP) to obtain the optimal $\delta p_t^Z$ values which ultimately incorporate the accepted orders and determine the zonal prices \cite{Euphemia}. 
\section{Analysis of the Input Data}
While the general framework of our model is applicable to any FBMC DAA market, it is worth noting that every single power market has its unique features \cite{DavisBook}, and thus any fundamental model needs to be adapted accordingly. Here we use the CWE DAA Power Market as case study, and for added clarity we describe its data structure before formulating the model. The CWE Power Market is composed of five zones of interest: Austria,
Belgium, Germany, France and the Netherlands. There are four major data sets: (i) the constraints data including the line-zonal PTDF and RAM \cite{JAOTool}, (ii) the power production data \cite{ENTSOEData}, (iii) the grid data \cite{ENTSOEGrid}, \cite{SciGrid}, and (iv) other time series data, typically recorded per zone, which also include their day-ahead estimates \cite{ENTSOEData}. Unless otherwise specified, the data is available at \cite{ENTSOEData}. We restrict our attention to the first $210$ days of 2019, which gives $|\mathcal{T}|=5040$ data points for each hourly time series.  We next describe data preprocessing in more detail and analyse the data briefly.
\subsection{Constraints Data}\label{consdata}
The data set, which we denote as $\mathcal{D}_\mathcal{LST}$, consists of hourly recorded $PTDF$, $RAM$, $F_{ref}$, $F_{max}$, and $FRM$ values, on transmission lines corresponding to each zone of interest and is available at \cite{JAOTool}. A brief description of the data features is given below:
\begin{enumerate}
    \item at each time, we observe data only a small subset of the critical branches $\mathcal{L}_{cb}$, corresponding to active constraints in the optimisation algorithm \cite{Euphemia}. 
    \item At each fixed time, for each critical branch, $l \in \mathcal{L}_{cb}$, we potentially observe multiple corresponding critical outages (CO), $s \in \mathcal{S}$. This gives the following data structure for each line-scenario-time combination $\mathcal{D}_{l,s,t}= \{PTDF, RAM, F^{ref}, F^{max}, FRM, FAV\}$. Since we do not observe all $(l,s,t)$ pairs for $l \in \mathcal{L}_{cb}$, $s \in \mathcal{S}$, and $t \in \mathcal{T}$, we denote the available ones as $\mathcal{LST}$, and $\mathcal{ST}:=\{(s,t): (l,s,t) \in \mathcal{LST}\}$.
    \item The chosen line orientation is not given in the data, and we account for this in our model,
    \item the constraint data is independent of the grid data, and thus we map this to the grid data described in subsection C, by matching the node names specified in the constraints data, with the ones in the grid data.
\end{enumerate}
We record $106$ CBs and $152$ COs in the considered time interval, and successfully map $102$ and $133$ respectively. This gives $|\mathcal{LS}|=336$ recorded line-scenario combinations.
\par
\begin{figure}[ht!]
    \centering
    \subfloat{{\includegraphics[width=3.9cm]{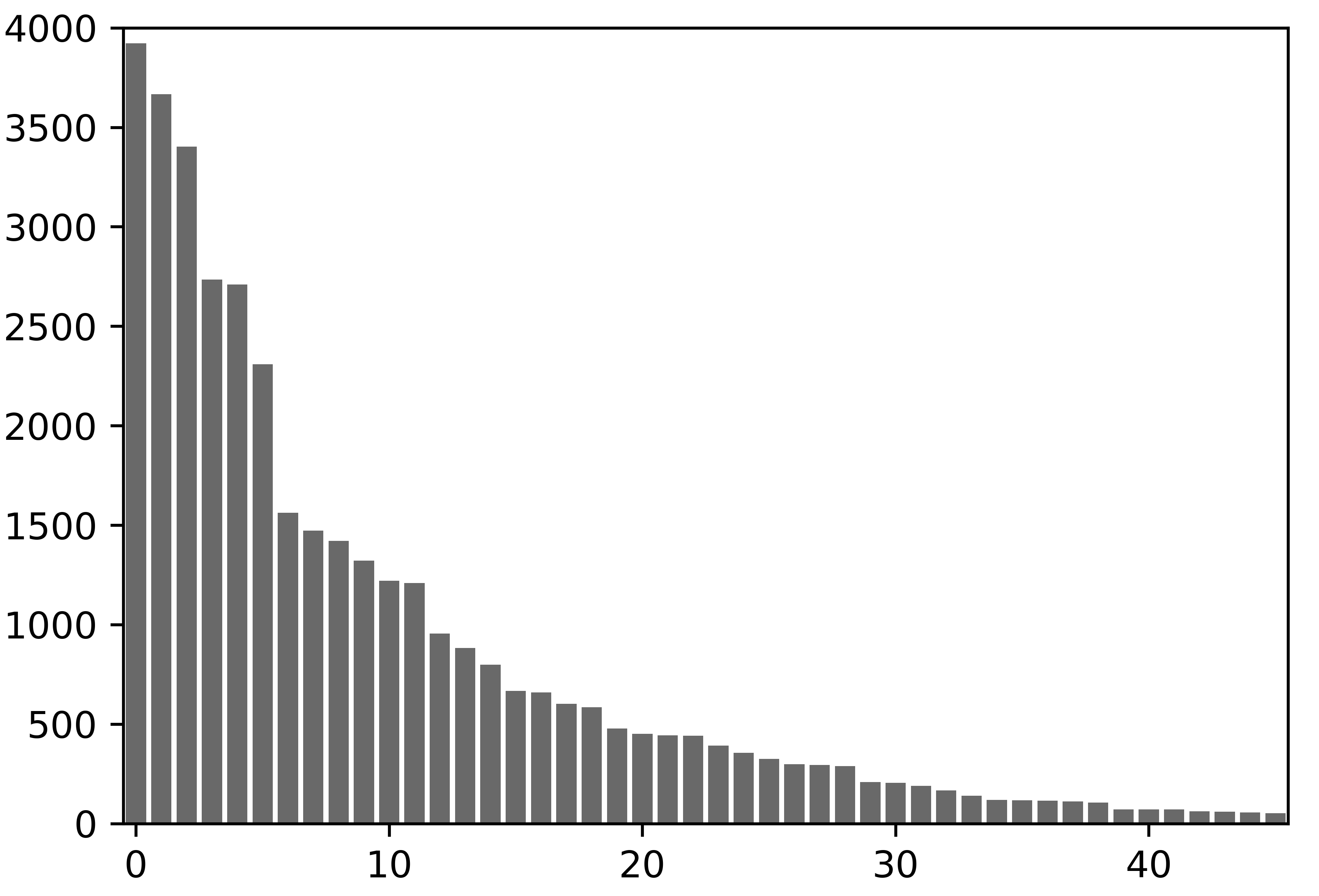} }}%
    \qquad
    \subfloat{{\includegraphics[width=3.9cm]{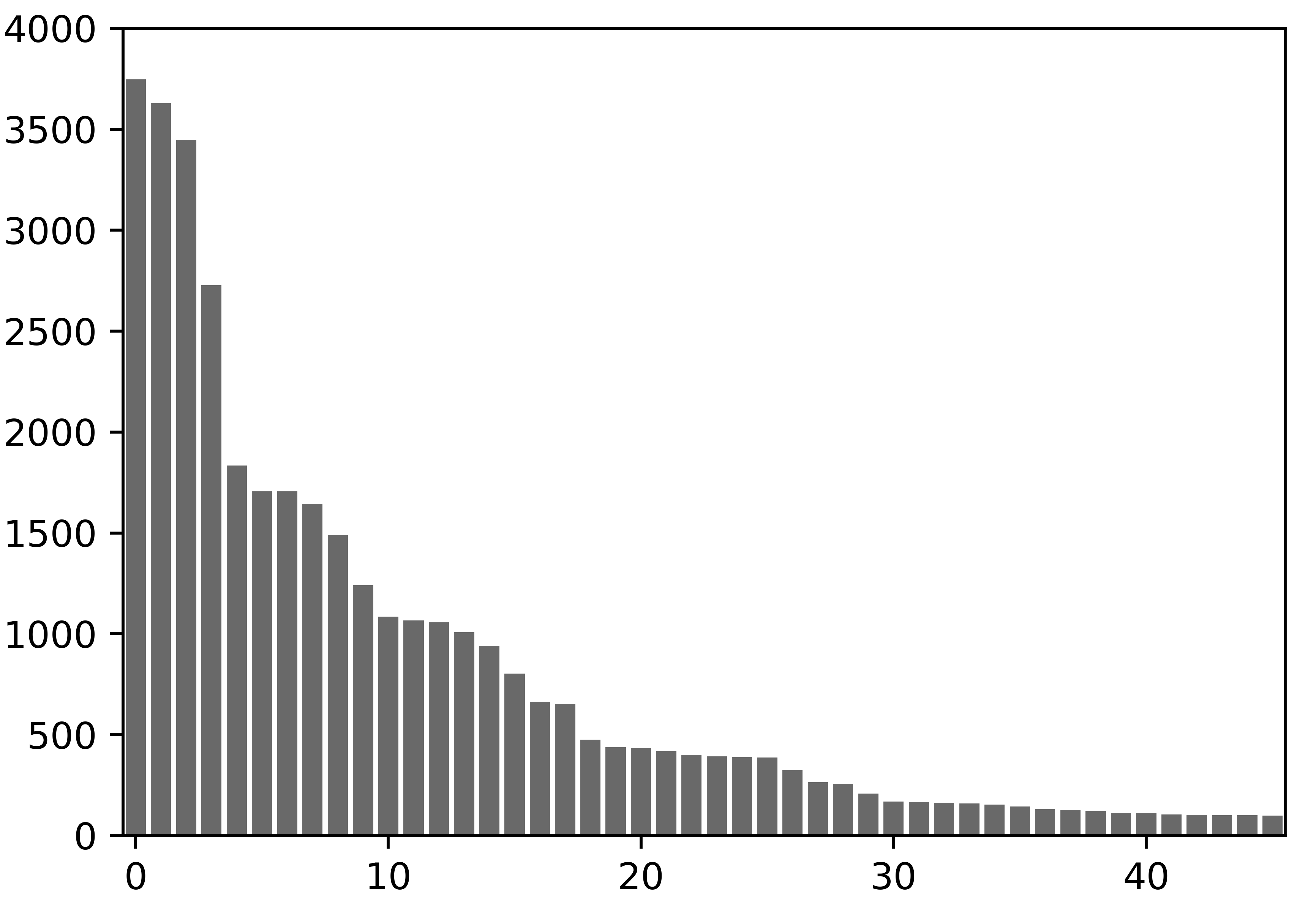} }}%
    \caption{Number of recorded data points for the first 45 lines (left) and scenarios (right), sorted in descending order of recorded information amount. }
    \label{fig:data_structure}
\end{figure}

The distributions of number of recorded data points for each $l$ or $(l,s)$ are not uniform but tends to resemble a power law, as observed in Figure \ref{fig:data_structure}. 
We thus observe that most of the information is recorded on a small fraction of lines as well as $(l,s)$ combinations. Adding the numbers in Figure \ref{fig:data_structure} and dividing by $|\mathcal{LS} ||\mathcal{T}|$, we see that the data amounts for only $2.3\%$ of all CB-CO (i.e $(l,s)$) combinations tracked by the TSOs. Clearly, the observed data is very spase and insufficient for many applications.
\subsection{Power Production Data}\label{powdata}
The generation units data can be found at \cite{ENTSOEData} and is comprised of hourly production levels for named power plants. We record 35 Austrian, 42 Belgian, 173 German, 142 French and 45 Dutch power plants. A data scraping algorithm was written to download the data. The data accounts for most of the non-renewable production in these countries, but does not include renewables. For each zone, we denote the plant set $\mathcal{GU}_z$, with $|\mathcal{GU}_z|=N_p^z$, the vector of production at each time by $\phi^z_t$ and the concatenation over zones as $\mathcal{GU}= \cup_{z\in \mathcal{Z}} \mathcal{GU}_z$, with $|\mathcal{GU}| = N_p$, and $\phi_t = ||_{z \in \mathcal{Z}}^V(\phi^z_t)$. Further $\Omega_N \in \mathbb{R}^{N_p \times N}$ is the canonical projection from all nodes to power plant nodes. Note that no geographical coordinates or any other indication of power plants' connection to the grid is given. Thus, we locate these mostly by hand since standard algorithms generally give unsatisfactory results.
\subsection{Grid Data}
\begin{figure}[ht!]
        \centering
        \includegraphics[width=8.7cm]{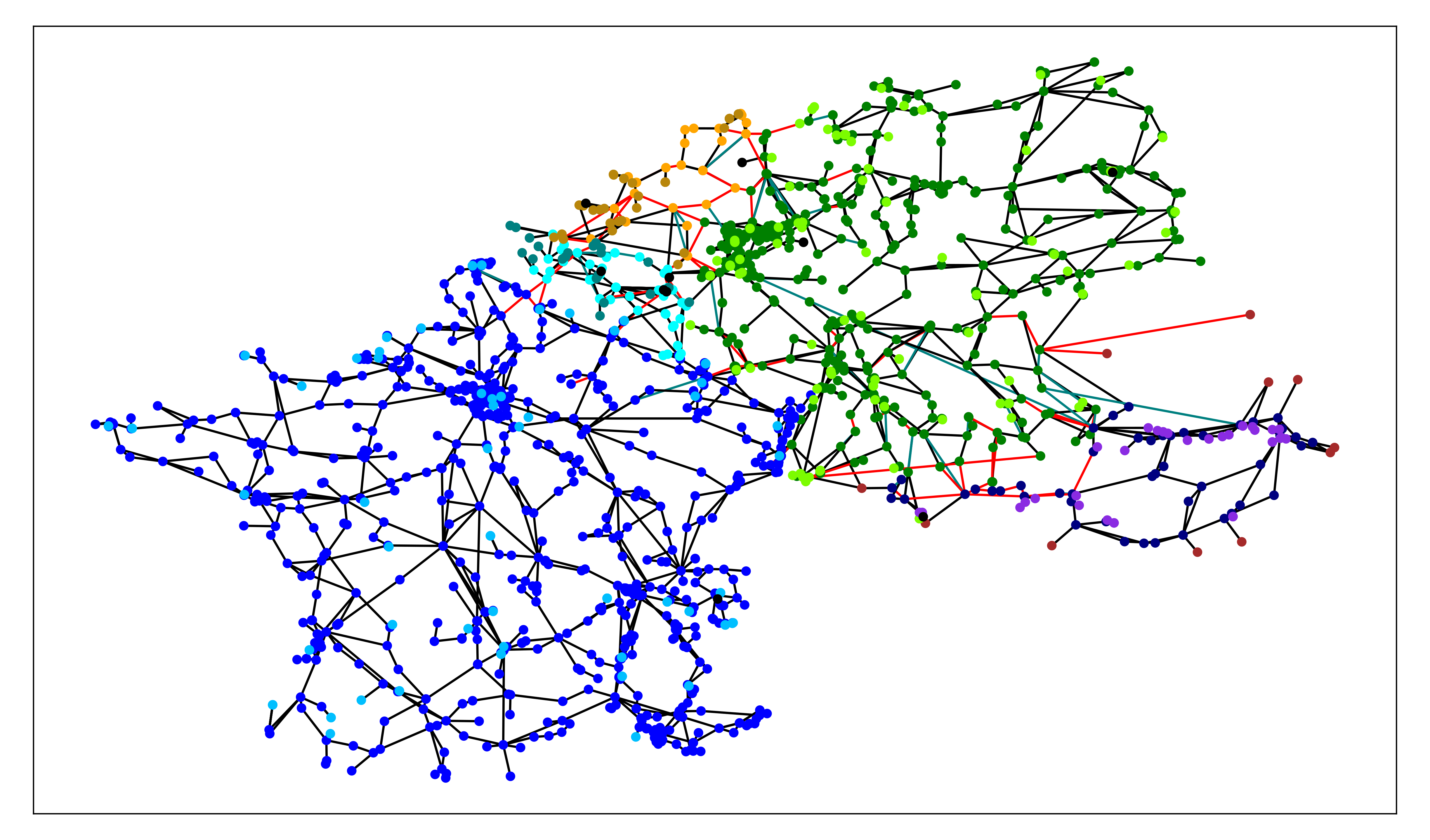}
        \caption{CWE Grid Model. Each country is assigned two colours, with the more frequent one representing consumption nodes, while the other represents power plants. Critical branches and outages are marked in red and cyan respectively. Lines that are both CB and CO are marked in red.}
        \label{fig:graph}
    \end{figure}

Most of the consumption grid nodes and edges are obtained by aggregating information from SciGRID \cite{SciGrid}, APG \cite{APG}, Elia \cite{Elia} and ENSTOE \cite{ENTSOEGrid}, where the latter is used
mostly for corrections by hand and visualisation purposes. We obtain the grid data as follows:
\begin{enumerate}
    \item SciGRID \cite{SciGrid} data is used as a starting point. This contains node and line information along with the geographical coordinates. Resistances ($R$), reactances ($X$), and susceptances ($b$) can be computed by using the given per-unit length constants, and as per the procedure described in \cite{SciGridUserguide}. We discard the data for Austria and Belgium due to incompleteness.
    \item We use the data available at \cite{APG} and \cite{Elia} to add the node and line information for Austria and Belgium. This requires geographical mapping. Since the Dutch grid from SciGRID is incomplete, we complete this by hand using \cite{ENTSOEGrid}. Any missing zone-connecting or critical lines are added by hand as per \cite{ENTSOEGrid}. Any lines with unknown length are assigned $1.1\cdot geod_l$, where $geod_l$ is the geodesic distance between the nodes connected by line $l$. The susceptances are then computed as per \cite{SciGridUserguide}. This gives the final consumption grid.
    \item Following the geographical locating described in subsection \ref{powdata} the power plants are connected to the closest node to yield the final CWE grid model.
\end{enumerate}
We denote the resulting susceptance vector by $b_0$. The graph structure is shown in Figure \ref{fig:graph}.
Note that the topology of $\mathcal{G} = (V,E)$ fully defines $A$, and the only physical characteristic needed are the line susceptances, $b$. To the best of our knowledge, this procedure makes use of the best available public information, but this will still introduce some error. Nevertheless, as we shall see in section IV, we only use this for initialisation and regularisation in our optimisation model, and so our model can tolerate inaccuracies in the Graph structure ($A$,$b$).
\subsection{Other time series data}
The rest of the data used in our model are: (i) zonal time series for the demand $\{D^z_t\}_{t \in \mathcal{T}}$, (ii) import and export giving the Net Export series $\{NEX^z_t\}_{t \in \mathcal{T}}$, (iii) solar $\{\phi^{s,z}_t\}_{t \in \mathcal{T}}$ and wind $\{\phi^{w,z}_t\}_{t \in \mathcal{T}}$ renewable energy production, and (iv) the zonal prices $\{S^z_t\}_{t \in \mathcal{T}}$, all of which can be found at \cite{ENTSOEData}. The day-ahead versions of $\{D^z_t\}_{t \in \mathcal{T}}$, $\{\phi^{s,z}_t\}_{t \in \mathcal{T}}$ and $\{\phi^{w,z}_t\}_{t \in \mathcal{T}}$ are also available at \cite{ENTSOEData}.
\section{Mathematical Optimisation Model}
The core idea of the model is to fit the GSKs, PAs and grid characteristics to the historical $PTDF$ and $F^{ref}$ data described in \ref{consdata}, via equations \eqref{lineZonalPTDF} and \eqref{Frefeq}. Direct fitting to the RAMs is not possible, but note that once $F^{ref}$ is recovered, the RAMs can be obtained as per \eqref{RAM eqs}. When $FRM$ and $FAV$ are unknown, a simple solution is to set them to the mean historical value or zero.
\par However, due to the sparsity of the data set $\mathcal{D}_\mathcal{LST}$, even for fixed $(b,A)$, multiple solutions exist for the GSKs and PAs. To simplify the analysis and increase data density, we assume that the GSK and base case are scenario independent, which is in agreement with literature proposals of defining the GSKs \cite{Schonheit}, \cite{Dierstein2017}, \cite{WPEN2015}. Thus we have that $G^1_{s,t} = G^1_t$ and $g^0_{s,t} =g^0_t$, which immediately implies that the $s$ dependence of the $\{PTDF_{s,t}\}_{(s,t) \in \mathcal{ST}}$ and $\{F^{ref}_{s,t}\}_{(s,t) \in \mathcal{ST}}$ data comes exclusively from $A^s$ and $b^s$. This reduces the number of time dependent variables and helps preventing over-fitting. However, the corresponding optimisation problem remains under-determined and thus we introduce a regularisation for $g^1_t$ in \ref{gskreg}. We then analyse the expressivity of $P_s(b,A)$ map, which is crucial to reconstruct the PTDFs in \ref{expresivity}, and we present the main optimisation model in \ref{mainmodel}.
\par
Before we proceed, we mention some observations and required adjustments. The GSKs are only defined on the production nodes, since the TSOs have virtually no control over the demand. Thus, we define $g^1_{z,t} \in \mathbb{R}^{N^z_p}$,  $\overline{g}^1_{z,t} \in \mathbb{R}^{N_p}$ and $G^1_{t}\in \mathbb{R}^{N_p \times Z}$ only over the production nodes. We write the equivalent of equation \eqref{powerlinearapprox} for production nodes only as
\begin{equation}\label{poweronlylinearapprox}
    \phi^z_t = g^0_{z,t} +\Delta \tilde{P}_{z,t} g^1_{z,t},
\end{equation}
where $g^0_{z,t} \in \mathbb{R}^{N^z_p}$ $\Delta \tilde{P}_z$ are also re-defined for production nodes only. We will frequently use equation \eqref{poweronlylinearapprox}, instead of \eqref{powerlinearapprox}, but this does not reduce the amount of available information, since no nodal demand data is available. Note that $\Delta \tilde{P}_z$ denotes the total power deviation from the base case production levels.
\subsection{A regularisation for the GSKs}\label{gskreg}
To obtain our desired GSKs regularisations, we use the JPW \cite{SchonheitAT} variant of the Sensitivity Coefficients Regression (SCR) model, which was proposed to define the GSKs in Austria \cite{SchonheitAT} and Germany \cite{Schonheit}. For each zone $z$ and time $\tau$, the model solves the linear regression problem denoted as $\mathcal{R}(z,p,\tau)$,
\begin{equation}\label{regressionproblem}
    \boldsymbol{\phi}^{p} = \theta + \mathbf{n}\eta + \mathbf{l}\lambda + \mathbf{a}\iota + \mathbf{R}\boldsymbol{\rho} + \mathbf{D}\boldsymbol{\delta} + \mathbf{X}\boldsymbol{\xi}+ \epsilon,
\end{equation}
for each plant $p$ in zone $z$, where $\epsilon$ is the error term, $\boldsymbol{\phi}^{p} \in \mathbb{R}^{T^{(\tau)}_p}$ is the vector of production levels for plant $p$, with $T^{(\tau)}_p=|\mathcal{T}^{(\tau)}_p|$ and $\mathcal{T}^{(\tau)}_p:= \{t: (\phi^z_t)_p \neq 0, t \in [\tau - w_z, \tau] \}$ being the index set of all data points for which plant $p$ has nonzero production and with times shifted backwards at most $w_z$ time units with respect to the time of interest, $\tau$.
\par 
The net position vector $\mathbf{n} \in \mathbb{R}^{T_p}$ can be expressed as $\mathbf{n}_t := \Delta P_{z,t}$, while $\mathbf{a}_t := S_t^{z}$ and $\mathbf{l}_t := D^{z}_{t}$ define the vector of zonal energy prices and demands respectively. Wind and solar production levels are considered as exogenous variables and each is assigned a column of $\mathbf{R}^{T_p \times 2}$. The dummy variable matrix $\mathbf{D}\in \mathbb{R}^{T_p \times 3}$ differentiates between peak/offpeak and summer/winter conditions, while  $\mathbf{X}\in \mathbb{R}^{T_p \times 3}$ is the interaction matrix of time dummy variables and net position change, such that $\mathbf{X}= diag(\mathbf{n})\mathbf{D}$. Finally $\theta$, $\eta$, $\lambda$, $\iota \in \mathbb{R}$ and $\boldsymbol{\rho} \in \mathbb{R}^2$, $\boldsymbol{\delta} \in \mathbb{R}^3$, $\boldsymbol{\xi} \in \mathbb{R}^3$ are the regression coefficients to be determined. 
\par
Since for each zone, the GSKs can be expressed as
\begin{equation}\label{gskdef2}
    (g^1_{z,t})_p = \frac{(\Delta \phi_t)_p}{\Delta P_{z,t}}; \hspace{6mm}\forall p \in \mathcal{GU}_z,
\end{equation}
where $\Delta \phi_t^z = \phi^z_t - g^0_{z,t}$ and $\Delta P_{z,t} = \sum_{p \in \mathcal{GU}_z} (\Delta \phi^z_t)_p$, the SCR model sets the GSK entries to be proportional to the infinitesimal version of \eqref{gskdef2} and thus \cite{Schonheit}
\begin{equation}
    \left(g^1_{z,\tau}\right)_p \sim \frac{\partial \phi_{p,\tau}}{\partial \Delta P_{z,\tau}} = \eta + \mathbf{D}_{\tau,:} \boldsymbol{\xi} =:\upsilon_{z,p,\tau}.
\end{equation}
The SCR then uses the fact that $1^Tg^1_{z,\tau}=1$ to obtain the GSK values by normalisation as 
\begin{equation}\label{gskbySCR}
\left(g^1_{z,\tau}\right)_p =\frac{\upsilon_{z,p,\tau}}{\sum_{p \in \mathcal{GU}_z}\upsilon_{z,p,\tau}}.
\end{equation}
Since data points for which $\left(\boldsymbol{\phi}^z_t\right)_p=0$ are excluded, the number of data points for some regression problems $\mathcal{R}(z,p,\tau)$ might not be sufficient. In this case, or if $\upsilon_{z,p,\tau}$ is below a (small) threshold value, the corresponding entry in $\upsilon_{z,p,\tau}$ (and thus in $g^1_{z,\tau}$) is set to zero \cite{Schonheit}.
\par
One can now apply the procedure for all $\tau \in \mathcal{T}$ to obtain the zonal GSKs which we denote as $G^1_z(\mathcal{T})$ and we denote this procedure as $SRC(z, \mathcal{T})$. It is suggested in \cite{SchonheitAT} that better results are obtained when using the day-ahead (D-1) version of the regression coefficients. For us, this poses a problem for the left hand side of \eqref{regressionproblem}, since the D-1 version of $\phi_t^z$ is $g^0_{z,t}$, which is unknown. However, we propose to achieve this by initialising $g^0_{z,t} \gets \phi_t^z$, obtaining $g^1_{z,t}$ via $SRC(z, \mathcal{T})$, updating $g^0_{z,t}$ via \eqref{poweronlylinearapprox} and then iterating again until we reach converge, which we measure by the norm of the difference of $g_{z.t}^1$ between two consecutive iterations. This gives us the GSK regularisation that we denote with a hat from here on, i.e. as $\hat{g}_{z,t}^1$, $\hat{g}_{t}^1$ and $\hat{G}_{t}^1$ respectively.
\begin{figure}[ht!]%
        \centering
        \subfloat{{\includegraphics[width=3.9cm]{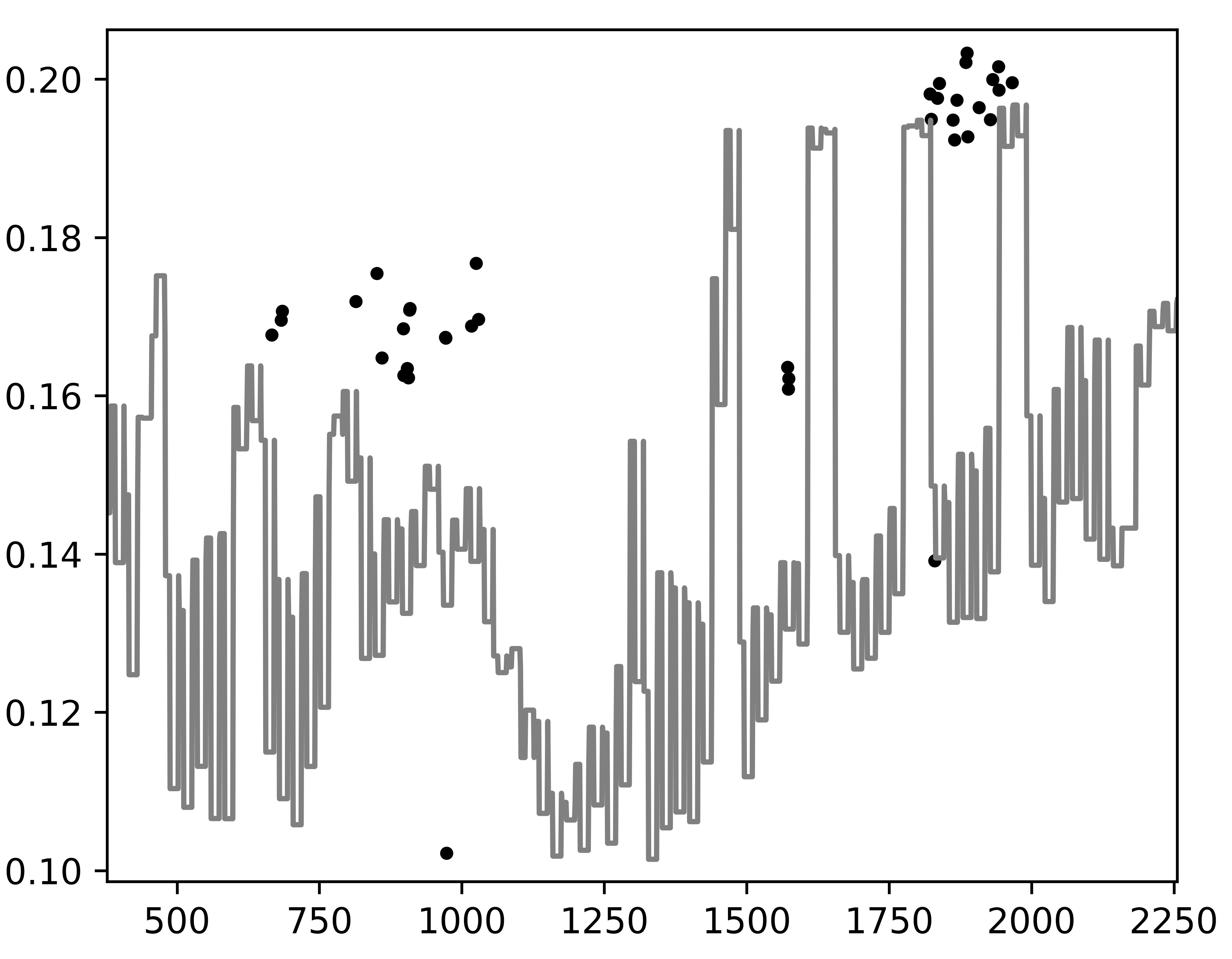} }}%
        \qquad
        \subfloat{{\includegraphics[width=3.9cm]{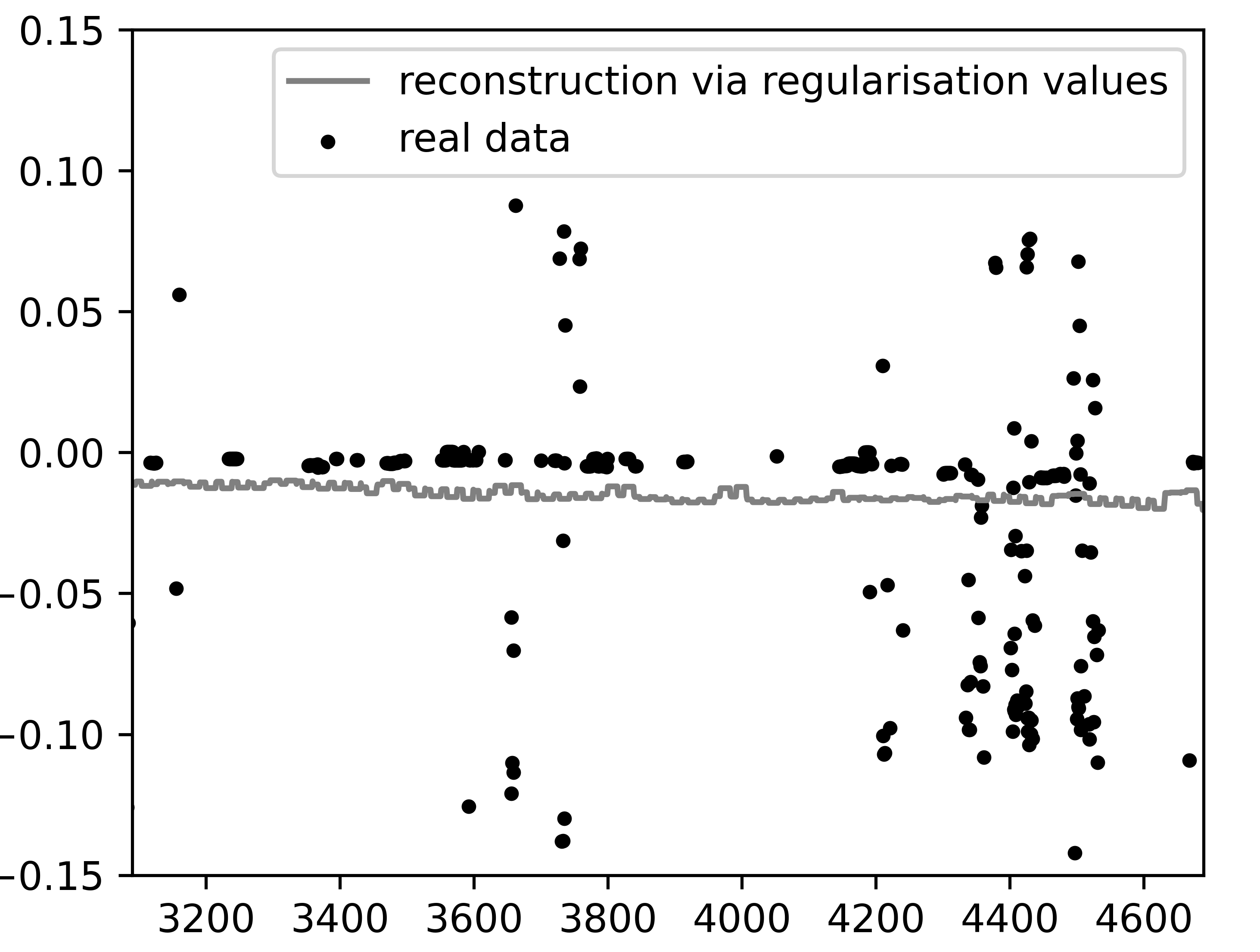} }}%
        \caption{PTDF values on an Austrian line (left) and a German-Dutch crossing line (right) showing the reconstruction of the PTDF based on the regularisation values of $b_0$ and $\{\hat{g}_{t}^1\}_{t \in \mathcal{T}}$. }%
        \label{fig:ptdfreconex}%
    \end{figure}
By using only the regularisation values $b_0$ and $\{\hat{g}_{t}^1\}_{t \in \mathcal{T}}$, we can already obtain a PTDF reconstruction via \eqref{lineZonalPTDF}. As seen in Figure \ref{fig:ptdfreconex}, this generally gives reasonable qualitative reconstruction suggesting that we obtained appropriate regularisations, since the PTDF data was not used up to this point. Nevertheless, the overall reconstruction error is rather large, and the regularisations' reconstruction performs poorly on many lines, particularly due to the inability to capture large variations, or be centred appropriately, as shown in the right part of Figure \ref{fig:ptdfreconex}. Further, no reasonable reconstruction of the RAMs is available at this point. This shows the need for our main optimisation model, introduced in \ref{mainmodel}.
\subsection{Expressivity of the Nodal PTDF map}\label{expresivity}
When running the main optimisation problem with $\Omega_N^T$, on many lines, we observe poor ability to reconstruct (expressivity) the PTDFs' large variation, regardless of the chosen GSK. One reason for poor expressivity is that some power plants cannot be differentiated by the grid structure. This is because by artificially connecting the power plants to the closest consumption node, multiple power plants inject power to the same single node, and thus are indistinguishable. We remedy this by replacing $\Omega_N^T$ with the map $M(x) \in \mathbb{R}^{N \times N_p}$ defined as
\begin{equation}
    M(x)_{:,p} = \sum_{p \in \mathcal{GU}} \Omega^p x_p,
\end{equation}
with $\Omega^p = \left[e_{v_1},e_{v_2},...e_{v_k} \right]\in \mathbb{R}^{N \times k}, \hspace{2.5mm} \forall v_i \in \mathcal{N}_k(p)$, where $e_q$ is the indicator vector for position $q$, $\mathcal{N}_k(p)$ is the set of $k$-nearest neighbours of plant $p$ (including itself), and $x: =||^V_{p \in \mathcal{GU}}x_p$. Thus, $x^p$ tells us how the power for plant $p$ is distributed across neighbouring nodes. By optimising over $x$, and enforcing that $0 \leq x_p \leq 1$ and $1^Tx =1$, we allow finding the most likely power injection pattern of each plant to the $k$ closest neighbours, instead of pre-defining a single fixed connection to the closest node, and by choosing a sufficiently small $k>1$ we can avoid over-fitting.
\subsection{The optimisation model}\label{mainmodel}
We now introduce the main optimisation model, explaining every term individually. The optimisation variables are $b$, $x$, the time series $\{g^1_t\}_{t \in \mathcal{T}}$, $\{\alpha_t^L\}_{t \in \mathcal{T}}$, and the constraints direction, which we model as binary variables $\nu_{l,s,t} \in \{-1,1\}$ for every $(l,s,t) \in \mathcal{LST}$. We introduce optimisation terms to (i) fit to the PTDF data, (ii) fit to the $F^{ref}$ data, (iii) regularise $b$, (iv) regularise $g_t^1$, and the corresponding physical constraints. We further regularise $\alpha_t^L$ implicitly by taking this to be constant over non-overlapping time windows of length $w_\alpha$, and $M(x)$ by choosing a small value of $k$. Note that we do not optimise over $A$ since this is NP hard, and because the number of variables would become very large, and thus we write $P_s(b) := P_s(b,A)$. This is reasonable since we expect to have a good initial estimate of the graph topology, but it is much less likely that we have accurate physical characteristics of this (such as $b$).
\par
We start by describing the PTDF term. Our choice of reference node might be mismatched by the TSO, and as a result we can recover $PTDF_{s,t}$ only up to a rank one matrix. This can be remedied by introducing the following mean centering: for each line $j$, let $\mathcal{ST}_j=\{(s,t):\,(j,s,t) \in \mathcal{LST}\}$, $\mathcal{L}_{s,t} := \{j: (j,s,t) \in \mathcal{LST} \}$ and let $Y \in \mathbb{R}^{L_{cb} \times Z}$ with
	 \begin{equation}\label{defY}
	 Y_{j,:}:=\frac{1}{|\mathcal{ST}_j|}\sum_{(s,t)\in \mathcal{ST}_j} (PTDF_{s,t})_{j,:},
	 \end{equation}
which is well defined since $\mathcal{ST}_j\neq\emptyset$ for all $j \in \mathcal{L}_{cb}$, and $PTDF_{s,t}  \in \mathbb{R}^{|\mathcal{L}_{s,t}|\times Z}$ is the concatenation of all the available PTDFs at $(s,t)$. Likewise, 
	 \begin{equation}\label{def_y}
	 \tilde{y}_j(b,s,t):=\frac{1}{N-1} (P_s(b))_{j,:}\mathbf{1}_N,
	 \end{equation}
where $\mathbf{1}_N \in \mathbb{R}^N$. The centred PTDF data is then
\begin{equation}
    PTDF^c_{s,t} = \nu_{s,t} \odot PTDF_{s,t} - \Omega_{s,t} Y,
\end{equation}
where $\odot$ denotes the element-wise product, $\Omega_{s,t} \in \mathbb{R}^{|\mathcal{L}_{s,t}|\times L_{cb}}$ is a projection map from the critical branches to the one observed for $(s,t)$. Similarly, the centred version of the nodal PTDF becomes
\begin{equation}
    P^c_s(b):=P_s(b) - \tilde{y}(b,s,t)\mathbf{1}_N^T.
\end{equation}
We further regularise $g^1_.$ by imposing constant value over a (small) time window $w_g$ via ${\theta_t^g} = \left \lfloor{t/w_g}\right \rfloor$. The PTDF optimisation term is then
\begin{equation}\label{ptdfterm}
    f_P(b,g^1(\mathcal{T}), x) = \sum_{s,t \in \mathcal{ST}}\| PTDF^c_{s,t} - \hat{\Omega}_{s,t}P^c_s(b) M(x) G^1_{\theta_t^g}\|_2^2,
\end{equation}
where $g^1(\mathcal{T})=\{g^1_{\theta_t^g}\}_{t \in \mathcal{T}}$, $\hat{\Omega}_{s,t} =\Omega_{s,t}\Omega_{cb}$. Note that $PTDF^c_{s,t}$ has to be computed just once, while $P^c_s(b)$ must be updated at every iteration for $b$. The PTDF reconstruction for the CBs can be then obtained for every $(s,t)$ as 
\begin{equation}
    \overline{PTDF}_{s,t} = P^c_s(b) M(x) G^1_{\theta_t^g} + Y.
\end{equation}
We can obtain the $F^{ref}$ optimisation term in a similar way. However, note that since $\alpha^L_.$ are centred around zero, $PSDF_s(b)$ centring is not necessary. Further, since the number of lines is very large, we only allow a small fraction of lines to have nonzero entries in $\alpha^L$. To achieve this we redefine $\alpha_L \in \mathbb{R}^{L_\alpha}$ to be the vector of PAs, and $\Omega_\alpha \in \mathbb{R}^{L \times L_\alpha}$ to be the projection map from the set of nonzero PAs to all lines. We further regularise $\alpha^L$ by imposing constant value over the time window $w_\alpha$ via $\theta_t^{\alpha} = \left \lfloor{t/w_\alpha}\right \rfloor$, to obtain
\begin{equation}\label{frefterm}
     f_F(b,\{\alpha^L_{\theta_t^{\alpha}}\}_{t \in \mathcal{T}}) := \sum_{s,t \in \mathcal{ST}}\|F^{ref,c}_{s,t} -\hat{\Omega}_{s,t}PSDF_s(b)\Omega_\alpha\alpha^L_{\theta_t^{\alpha}}\|^2_2,
\end{equation}
where $F^{ref,c}_{s,t}= \nu_{s,t} \odot F^{ref}_{s,t}-\Omega_{s,t}\mu(F^{ref})$, with $\mu(F^{ref})$ being the mean reference flow values over the critical lines. By comparing \eqref{frefterm} to \eqref{Frefeq}, one can observe that we neglected the term given by $g_t^0$, which can be written as a function of only $g^1_t$ via \eqref{poweronlylinearapprox}. In numerical experiments we found that this term tends to only introduce noise, without improving the quality of fit or generalisation error. Since the effect of $\phi_t$ is taken into account by mean centring, this suggests that the GSKs do not play a role in the definition of the base case, which is to be expected. We also enforce that $- \pi/6 \leq\alpha^L_{\theta_t^{\alpha}} \leq \pi/6$ as per \cite{WPEN2014}, and that the reference flows do not exceed maximum allowed flows on the critical lines 
\begin{equation}
  \overline{lb} \leq  P^{cb}_{\mathcal{S}}(b) \alpha^L_{\theta_t^{\alpha}} \leq \overline{ub},
\end{equation}
for every $\theta_t^{\alpha}$, where $ P^{cb}_{\mathcal{S}} = ||^V_{s \in \mathcal{S}}\Omega_{cb}P_s(b)$, the flow bounds are $\overline{ub} = ||^V_{s \in \mathcal{S}}\left(F^{max} - \mu (F^{ref})\right)$, and $\overline{lb} = ||^V_{s \in \mathcal{S}}\left(-F^{max} - \mu (F^{ref})\right)$, with $F^{max}$ being the vector of maximum recorded $F^{max}_{s,t}$ values for each line.
\par 
The regularisation terms can simply be written as
\begin{equation}\label{gsk_and_b_reg}
    f_G(\{g^1_{\theta_t^g}\}_{t \in \mathcal{T}}) = \sum_{t \in \mathcal{T}} \|\hat{g}_t^1 - g_{\theta_t^g}^1 \|^2_2, \hspace{3mm}   f_b(b) = \|b_0 -b\|_2^2.
\end{equation}
By collecting all the terms, the final optimisation problem is:
\begin{align*}\label{mainoptpb}
    \min_{\omega}& f_P(b,\{g^1_{\theta_t^g}\}_{t \in \mathcal{T}}, x)  + \lambda_F f_F(b,\{\alpha^L_{\theta_t^{\alpha}}\}_{t \in \mathcal{T}})+ \\&\lambda_{G} f_G(\{g^1_{\theta_t^g}\}_{t \in \mathcal{T}})+ \lambda_b f_b(b,)\\
    \text{s.t. } &b \geq \delta_b, \text{ and } \nu_{s,t} \in \{-1,1\}^{|\mathcal{L}_{s,t}|} \\
    \text{}&1^Tg^1_{z,\theta_t^g} = 1 ,\text{ and }\text{} -\delta_g^z\leq g^1_{z,\theta_t^g} \leq \delta_g^z ,\text{ } \forall z \in \mathcal{Z}, \text{}t \in \mathcal{T},\\
    \text{}& 1^Tx^p=1, \text{ and }\text{} 0\leq x^p \leq 1, \text{ }\forall p \in \mathcal{GU},\\
    &\overline{lb} \leq  P^{cb}_{\mathcal{S}}(b) \alpha^L_{\theta_t^{\alpha}} \leq \overline{ub}, \text{ and }-\frac{\pi}{6} \leq\alpha^L_{\theta_t^{\alpha}} \leq \frac{\pi}{6}, \forall t \in \mathcal{T},
\end{align*}
where $\delta_b>0$ is a small number, $0<\delta^z_g\leq 1$ is the maximum allowed GSK value per entry, $\lambda_. \in \mathbb{R}_+$ are regularisation parameters, and the full variable set is denoted as $\omega = (b,\{g^1_{\theta_t^g}, \alpha_{\theta_t^{\alpha}}^L\}_{t \in \mathcal{T}}, x, \nu_{s,t \in \mathcal{ST}})$. Note that $b$ is fixed in principle, and once found it can be directly used for future predictions, but $g_{\theta_t^g}^1$ and $\alpha_{\theta_t^{\alpha}}^L$ being time dependent can not. However, these can be used for statistical analysis or forecasting. 
\section{Numerical Optimisation}
The optimisation problem is highly nonlinear and non-convex over the variable set $\omega$, and very large, with about $4$ million variables, but the vast majority of variables are the $g^1_{\theta_t^g}$, which are decoupled in time. Thus, we propose an alternating minimisation approach, by solving for one variable at a time, while keeping the others fixed, and using the most recent updates. This has the following steps: (i) solve for each $\nu_{s,t}$ independently, (ii) solve for $b$ via stochastic optimisation, (iii) solve for $x$ via stochastic optimisation, (iv) solve for $\{g^1_{\theta_t^g}\}_{t \in \mathcal{T}}$, for each time independently, via Convex Quadratic Programming (CQP), (v) solve for $\{\alpha^L_{\theta_{\alpha}{(t)}}\}_{t \in \mathcal{T}}$, for each time independently, via CQP, and (vi) iterate until convergence or time limit is reached. This approach does not guarantee the algorithm reaches the global minimum, since the resulting problems for $b$ and $\nu_{s,t}$ are not marginally convex. However, it makes the problem computationally feasible and leverages the marginally convex structure of the objective function in $g^1_t$ and $\alpha^L_{\theta_{\alpha}{t}}$ as well as time decoupling. Note that we are guaranteed to reduce the objective function value at every optimising step for $\nu_{s,t}$, $g_{\theta_t^g}^1$, and $\alpha_{\theta_t^{\alpha}}^L$, but not for $b$ and $x$, and thus, great care needs to be taken when optimising over these. We next describe in detail optimising over each set of variables.
\subsection{Optimising over $\nu_{s,t}$}
We optimise over each entry $\nu_{l,s,t} \in \{-1,1\}$ such that $(l,s,t) \in \mathcal{LST}$ independently, by choosing the value that decreases the objective function value most. With all the other quantities pre-computed, this is fast since we only have to compare two cases for each variable $\nu_{s,t}$. Further, since the data points are independent, this approach gives the global optimal solution when $b$, $\{g^1_{\theta_t^g}, \alpha_{\theta_t^{\alpha}}^L\}_{t \in \mathcal{T}}$ and $x$ are fixed.
\subsection{Optimising over b}
Due to the large number of scenarios $|\mathcal{S}|$, it is extremely expensive to compute all the $P_s(b)$ values, since these require the computation of $(\tilde{A}^T_s\tilde{B}_s\tilde{A}_s)^{-1}$ for every $s$ at every iteration. Further, the number of times $|\mathcal{T}|$ is also very large, making the computation of $f_P(b)$, $f_F(b)$, $\nabla_b f_F(b)$ and $\nabla_b f_P(b)$ extremely time and memory intensive, and thus we use tensor implementation in $PyTorch$. Computing the Hessian is impractical, and thus we resort to first order methods. We use stochastic optimisation, at each iteration $k$, by randomly choosing only a subset $\mathcal{T}_k$ of $\mathcal{T}$ (but all corresponding $s$), because this: (i) dramatically reduces the energy usage per iteration, due to much fewer inversions and $P_s(b)M(x)G^1_{\theta_t^g}$ products, (ii) it reduces the computational time, and (iii) we observe much larger objective function decrease per iteration when compared to the deterministic optimisation step, since the landscape of $f_P(b)$ is noisy, and stochastic methods act as a smoother. To allow step size adjustment, we use ADAM \cite{ADAM} as optimiser, and solve for $K_b$ iterations in every cycle.
\subsection{Optimising over x}
Although the problem for $x$ is convex, and we could solve via CQP, the double sum over $s$ and $t$ makes the problem excessively large, yielding infeasible memory requirements when using $Mosek$. Thus, we perform stochastic optimisation using the ADAM optimiser, similarly to the optimisation over $b$. This makes the computations less memory and energy intensive, but does not guarantee convergence to the optimal solution $x^*(b,G^1)$, even when all the other variables are fixed, since ADAM may not converge even on convex problems \cite{AMSGrad}. However, we find this to be a good trade-off between computational feasibility and making optimisation progress.
\subsection{Optimising for $g^1_{\theta_t^g}$}\label{optovergsk}
We solve for each $g^1_{\theta_t^g}$ independently since the problem is decoupled in time. The resulting optimisation problem reads
\begin{align*}
    \min_{g^1_\tau}&\sum_{\mathcal{F}_{\tau}^g}\| PTDF^c_{s,t} - \hat{\Omega}_{s,t}P^c_s(b) M(x) G^1_{\theta_t^g}\|_2^2+ \\ &\hspace{8mm}\lambda_{G}\|\hat{g}_t^1 - g_{\theta_t^g}^1 \|_2^2\\
    \text{s.t. }&1^TG^1_{\theta_t^g} = 1 ,\text{ } \forall t \in \mathcal{T},\hspace{3mm}
    \text{ }0\leq (G^1_{\theta_t^g})_{ij} \leq 1, \text{ }\forall t \in \mathcal{T}&,
\end{align*}
where $\mathcal{F}_{\tau}^g:= \{(s,t):s \in \mathcal{S}, \theta_t^g=\tau\}$. The problem is convex, and we solve it using $Mosek$'s API in $Python$, but other convex optimisation solvers could be used. 
\subsection{Optimising over $\alpha^L_{\theta_t^{\alpha}}$}\label{optoveralpha}
Similarly to \ref{optovergsk}, the problem is convex, and decoupled in time for every $\theta_t^{\alpha}$. This reads 
\begin{align*}
    \min_{\alpha_{\tau}^L}& \sum_{s \in \mathcal{S}, \{t: \theta_t^{\alpha} = \tau\}}\|F^{ref,c}_{s,t} -PSDF_s(b)\Omega_\alpha\alpha^L_{\theta_t^{\alpha}}\|^2_2\\
    \text{s.t. }&\overline{lb} \leq  P^{cb}_{\mathcal{S}}(b) \alpha^L_{\theta_t^{\alpha}} \leq \overline{ub}, \text{ } - \frac{\pi}{6} \leq\alpha^L_{\theta_t^{\alpha}} \leq \frac{\pi}{6}, \text{ } \forall t \in \mathcal{T}
\end{align*}
and we solve it using $Mosek$'s API in $Python$.
\subsection{Obtaining new values for $g^1_{\theta_t^g}$ and $\alpha^L_{\theta_t^{\alpha}}$ after the model fit}
Once the solution $\omega$ of the main optimisation model is available, one can obtain new values for $g^1_{\theta_t^g}$ and $\alpha^L_{\theta_t^{\alpha}}$ for $t \notin \mathcal{T}$ ($t > \max_{\tau\in \mathcal{T}}\tau$) by pre-fitting $\nu_{s,t}$ and then solving the corresponding convex optimisation problems \ref{optovergsk} and \ref{optoveralpha} once. The number of variables is reduced dramatically and once the $P_s(b)$ maps were pre-computed, this problem can be solved on a usual personal computer in at most few of minutes.
\section{Numerical Results}
We evaluate the model for: (i) in sample quality of fit, (ii) generalisation for $(l,s,t) \notin \mathcal{LST}$, with $l \in \mathcal{L}_{cb}$, $s \in \mathcal{S}$, $t \in \mathcal{T}$, i.e. for unobserved $(l,s,t)$ containing observed individual elements, and (iii) generalisation for unobserved $(l,s,t)$ with new times $t \notin \mathcal{T}$ and observed $l \in \mathcal{L}_{cb}$ and $s \in \mathcal{S}$. The first evaluation requires in-sample tests while for the other two we perform out of sample tests. Due to the data sparsity, it is impossible to split the data in a train and test set to perform out-of-sample tests for (ii) and (iii) simultaneously. Thus, we split the data twice, fit the model to each training set, and perform tests on the corresponding test set.
\par
For the first train-test split, we cannot simply remove (say) $20\%$ of the data at random and add it to the test set, because this leaves many $(l,s)$ combinations and $t$ indices with no data. To resolve this, we propose the following heuristic: (i) select an available $(l,s,t)$ point at random, (ii) if the number of points on $(l,s)$ and at time $t$ are above the thresholds $m_{LS}$ and $m_T$ respectively, remove the point from the train set, and add it to the test set, (iii) otherwise do not remove the point but mark it such that it cannot be chosen again, (iv) repeat until the desired test set size is achieved or no more points can be removed. We choose $m_T =8$, since this is the average data density with respect to time, and allowing a smaller value would make data rich time indices become data scarce. We choose $m_{LS} = 50$, but generally it is $m_T$ that has a strong impact on maximum fraction of data that can be removed. In this case, the maximum fraction of removed points is about $18\%$, but this corresponds to is highly inter-dependent choices. Thus, we choose to assign $85\%$ and $15\%$ to the train ($\mathcal{D}_1^{train}$), and test ($\mathcal{D}_1^{test})$ set respectively. We fit the model to the train set and compute the out-of-sample error on the test set.
\begin{figure*}[t]
    \centering
    \includegraphics[width=0.935\textwidth]{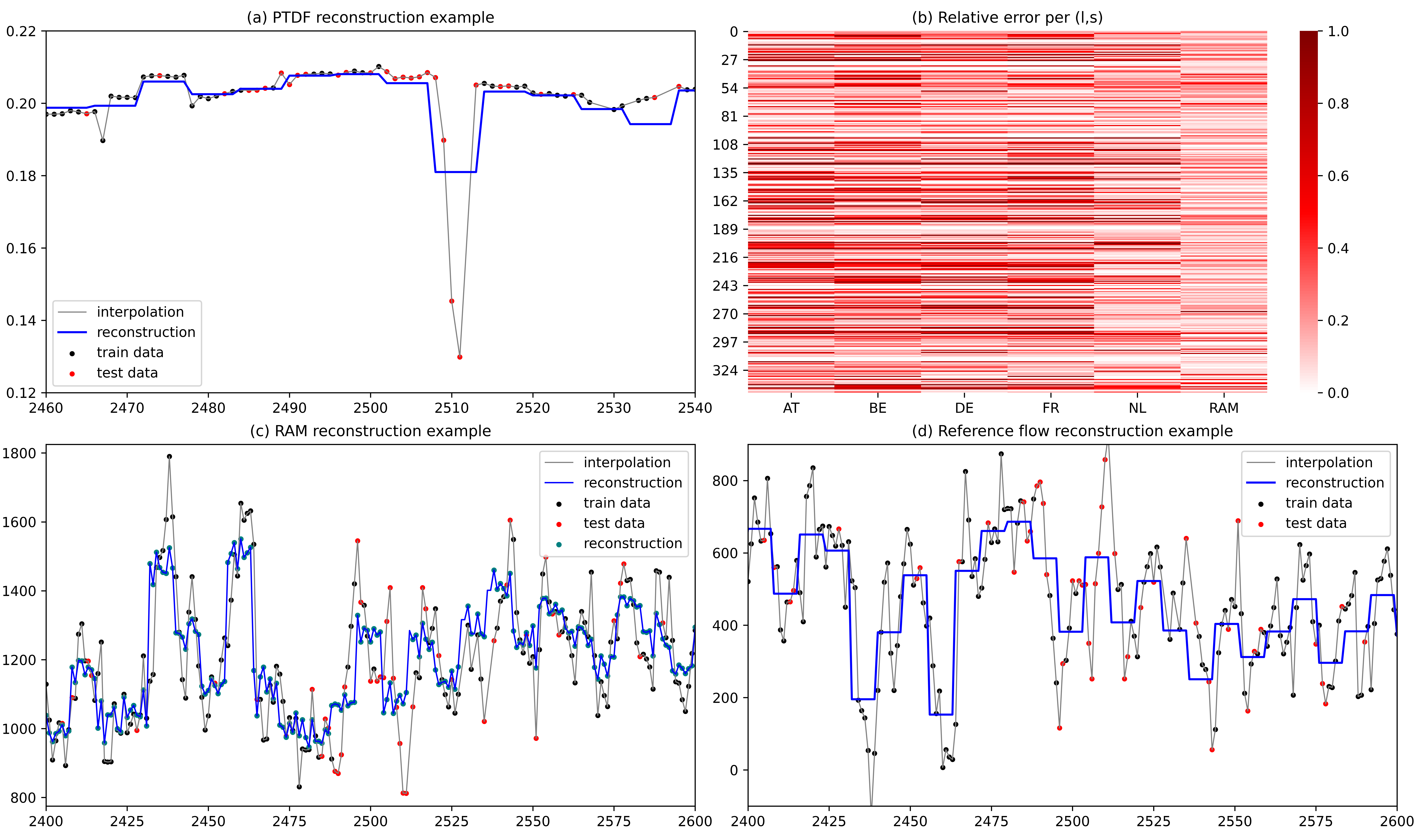}
    \caption{Detailed reconstruction results. Plots (a), (c) and (d) show the PTDF (French entry), RAM, and $F^{ref}$ reconstruction results for a representative (French-German) line. The blue lines (and dots) indicate the reconstruction, while the black and red dots indicate the training and test data points respectively. The grey line is an interpolation between available data points for better visualisation only. The heat map in (b) indicates the relative RAM error in the sixth column, and relative PTDF error for the five countries, corresponding to the first five columns. }
    \label{fig:linereconstructionexample}
\end{figure*}
\par
The second train and test sets are obtained by splitting the time indices at random, such that $20\%$ of the times are assigned to $\mathcal{D}_2^{test-t}$ and the rest to $\mathcal{D}_2^{train}$. Further, $\mathcal{D}_2^{test-t}$ is randomly split in $\mathcal{D}_2^{test-fit}$ and $\mathcal{D}_2^{test-lst}$ which receive $75\%$ and $25\%$ respectively. We fit the model only by using $\mathcal{D}_2^{train}$. However, to test the model we need GSKs, PAs and $\nu_{s,t}$ for each time interval, but these are unavailable. We obtain these by using fixed $b$, $x$ (fit based on $\mathcal{D}_2^{train}$), and fit for $g_{\theta_t^g}^1$, $\alpha^L_{\theta_t^{\alpha}}$ and $\nu_{s,t}$ only once on $\mathcal{D}_2^{test-fit}$. Out of sample tests are performed on $\mathcal{D}_2^{test-lst}$. This is an indication of generalisation quality to new times, given the time-independent variables $b$ and $x$, and thus the corresponding test evaluates if they give a good representation of the real grid.
\par 
We run the algorithm on a $72$ core machine with $768$ GB of memory. We find choosing $k=5$, $w_\alpha = 8$ and $w_g = 6$ to be a good trade-off between quality of fit and generalisation ability. We stop after running the algorithm for $14$ complete cycles, because the objective function value decrease per cycle becomes rather small, and the computational cost is high. 
\subsection{PTDFs and RAMs reconstruction}
To evaluate the quality of our reconstruction, we use multiple error measures. First, we use the average absolute mean error which we denote as $\delta_{abs}(.)$. To estimate the quality of fit, we consider $\delta_{\sigma}(.):=\delta_{abs}(.)/ \sigma (.)$, where $\sigma (.)$ is the standard deviation of the data set of interest. We use this ratio because if deviations are large, then it is expected that the ability to fit reduces, and thus we should penalise the model less. Further, we are interested in the performance of the model, and since to our knowledge, our model is the first to reconstruct the PTDFs and RAMs, we can only compare to a benchmark model. The benchmark model assumes that all data variation is noise, and thus it reconstructs the data by simply taking the empirical mean value for every line, and assigns this to every single $(s,t)$ combination with $s \in \mathcal{S}$ and $t \in \mathcal{T}$. We define the benchmark model error as $\delta_{null}(.)$ and the corresponding relative error with respect to the benchmark model as $\delta_{rNull}(.)=\delta_{abs}(.)/\delta_{null}(.)$. Finally, if one is to use these constraints in the optimisation problem for price finding \cite{Euphemia}, the relative error with respect to the mean value is more relevant, as this measures the fractional error in our constraints. We denote this as $\delta_{\mu}(.):= \delta_{abs}/\mu(.)$, where $\mu(.)$ takes the mean of the corresponding data set. However, since the PTDFs are mean zero centred we define $\mu(.)$ to be the mean of absolute values in this case.
\par
\begin{figure*}[ht!]
  \centering
  \includegraphics[width=\textwidth, height =13cm]{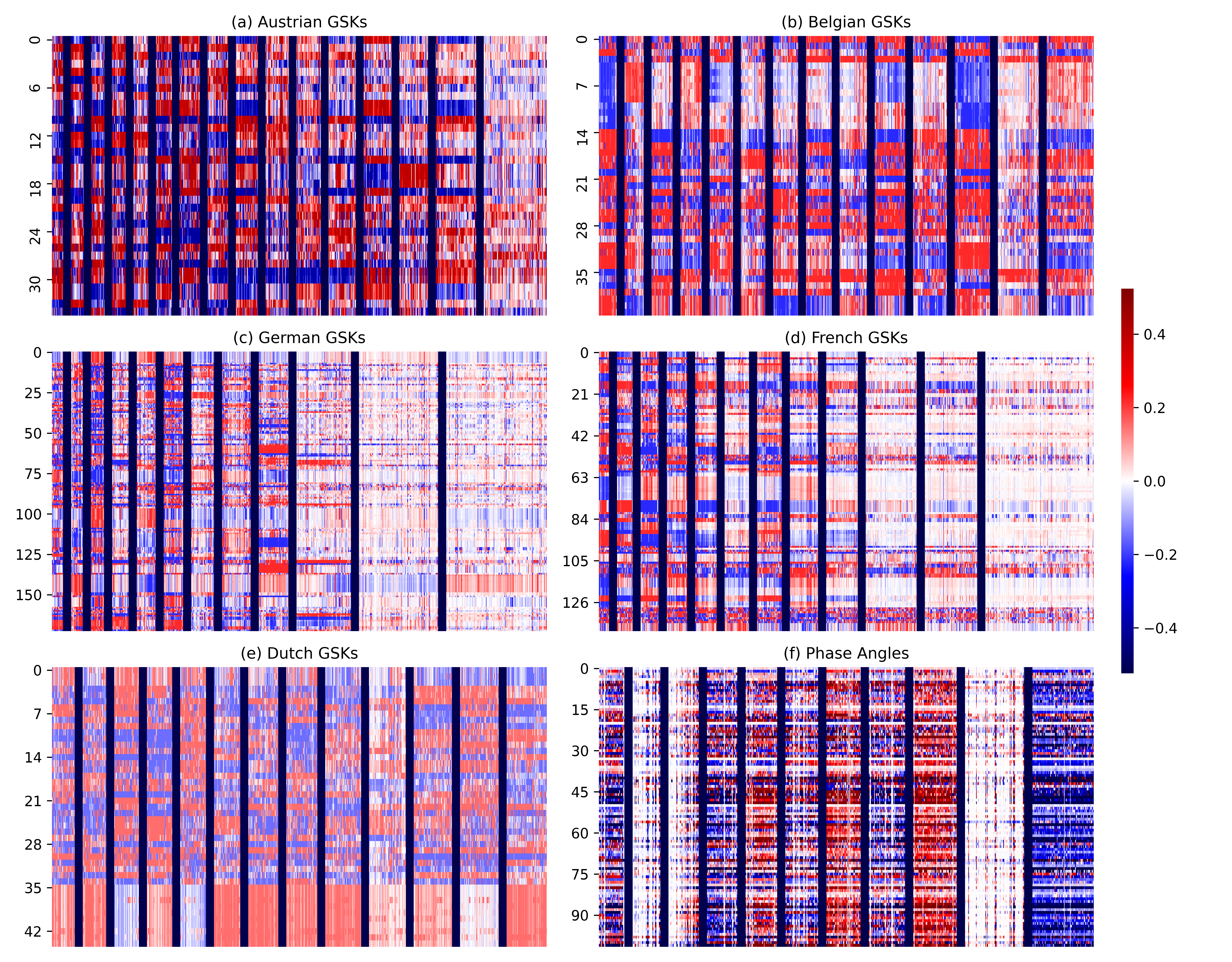}
  \caption{Clustering results for the Generation Shift Keys and Phase Angles. For each subplot, the $x$ axis corresponds to time indices while the $y$ axis corresponds to entries in $g^1_t$ and $\alpha_t^L$ respectively, and the vertical dark bars are delimiters between the clusters. }
  \label{fig:clusteringresults}
\end{figure*}
We show the overall errors obtained across the whole data set in Table \ref{tab:resultstable}. Considering that only about $2.3\%$ of the existing data is available, and thus used as input in our model, we obtain good reconstruction quality. Since $\delta_{\mu}$ is rather small, the reconstructed PTDFs and RAMs can be used to reconstruct the feasibility domain at each time step or for forecasting with reasonable error.  Further, we obtain good generalisation for both $\mathcal{D}_1$ and $\mathcal{D}_2$ tests, suggesting that our model is performing well for constraints reconstruction. The slightly better generalisation for $\mathcal{D}_1$ when compared to $\mathcal{D}_2$ could be because the grid changes, albeit slowly, over time. 
\begin{table}[ht!]
\begin{center}
\begin{tabular}{ | c || c| c |c| c| } 
\hline
Error measure& $\delta_{abs}$ & $\delta_{\sigma}$ & $\delta_{rNull}$ & $\delta_{\mu}$ \\ 
\hline
\hline
PTDF on $\mathcal{D}^{train}_1$  & 0.0122 & 14.4\% & 36.1\% & 22.7\% \\ 
\hline
PTDF on $\mathcal{D}^{test}_1$ & 0.0119 & 14.0\%  &35.2\%& 22.1\%\\ 
\hline
$F^{ref}$ on $\mathcal{D}^{train}_1$  & 88.1 & 36.1\%  & 39.6\%& 11.7\% \\ 
\hline
$F^{ref}$ on $\mathcal{D}^{test}_1$ & 96.6 & 39.5\%  &43.4\%& 12.8\%\\
\hline
RAM on $\mathcal{D}^{train}_1$  & 93.1 & 42.6\% & 54.9\%  & 16.0\%\\ 
\hline
RAM on $\mathcal{D}^{test}_1$ & 94.2 & 43.1\% &55.6\% & 16.2\%\\
\hline
\hline
PTDF on $\mathcal{D}^{train}_2$  & 0.012 & 14.2\% & 35.5\% & 22.3\% \\ 
\hline
PTDF on $\mathcal{D}^{test-lst}_2$ & 0.014 & 16.1\%  &40.5\%& 25.4\%\\ 
\hline
$F^{ref}$ on $\mathcal{D}^{train}_2$  & 87.5 & 35.8\%  & 39.3\%& 11.6\% \\
\hline
$F^{ref}$ on $\mathcal{D}^{test-lst}_2$ & 130.4 & 53.4\%  &58.5\%& 17.3\%\\
\hline
RAM on $\mathcal{D}^{train}_2$  & 92.7 & 42.4\% & 54.7\%  & 15.9\%\\ 
\hline
RAM on $\mathcal{D}^{test-lst}_2$ & 105.1 & 48.2\% &62.1\% & 18.1\%\\
\hline
\end{tabular}
\newline
\caption{\label{tab:resultstable} Train and test errors for the two data splits $\mathcal{D}_1$ and $\mathcal{D}_2$.}
\end{center}
\end{table}
Grid elements are added, removed or changed, and thus one might consider to periodically re-calibrate $b$ to new data, starting with the old $b$ in the optimisation process. By analysing $\delta_{\sigma}$, one can see that we obtain a better quality of fit to the PTDFs when compared to the RAMs. This is expected, because the PTDFs are simpler to compute, involving fewer variables and simpler physical equations. Finally, our model is clearly superior to the benchmark model, and thus is superior to the naive approach of treating data on every line as purely random. Therefore we capture meaningful structure in the data, and we further illustrate this
by showing more detailed results in Figure \ref{fig:linereconstructionexample}. 
\par
We can see that the reconstruction fits the training signal well, and implies reasonable values for time indices where data is unavailable. By analysing Figure \ref{fig:linereconstructionexample} (a), one can observe that we partially capture the sharp change in the PTDF although these points were not part of the training set. We also obtain good reconstruction for the $F^{ref}$ and RAM, although again we cannot fully capture sharp variations. This is because of the introduced time window values chosen for $w_{\alpha}$ and $w_g$. We can confirm that much better fit is obtained by letting these to one, but in this case the out of sample generalisation is rather poor. Thus, we found $w_{\alpha}=8$ and $w_g=6$ to be a good trade-off.  Note that by \eqref{RAM eqs}, even with fully available $F^{ref}$ values, the RAM cannot be reconstructed without knowing the other terms. To obtain full RAM reconstruction, for each $(l,s)$ pair, we thus always use the last available values at each time index. For this reason, we mark the points where exact $F^{max}$, $FRM$ and $FAV$ are used by blue dots in Figure \ref{fig:linereconstructionexample} (c). Finally, Figure \ref{fig:linereconstructionexample} (b) shows the $\delta_{\mu}$ value, computed for each available line scenario combination, that is each $(l,s) \in \mathcal{LS}$. We can observe that our model generally performs well, although some error entries are rather large. This could be because of small amount of information on some lines, but also because of imperfections in the grid structure, particularly regions where fine grid data is unavailable. While the optimisation model proposed in \ref{mainmodel} dramatically improves the quality of fit, due to the problem structure, multiple solutions exist, and finding the global minimum is NP hard. Further, we do not optimise over $A$, as this is also NP hard and would render the problem computationally infeasible. Nevertheless, our results generally give good reconstruction and generalisation, and outperform the naive approach. Further, we next show that we recover structure in the GSKs and PAs.
\subsection{Structure Analysis of reconstructed GSKs and PAs}
We perform clustering analysis to reveal groups of similar zonal GSKs and PAs respectively. We cluster at zonal level because the TSO in each zone chooses the GSKs independently of the others. We use the k-means algorithm \cite{Kmeans}, which in this case we found to give better results when compared to the Gaussian Mixture Models \cite{GMM} solved with via Expectation Maximisation. We run the algorithm 50 times and choose the result that gives the smallest within cluster sum of squares. We choose the number of clusters by using the elbow method, but since there is no clear steep change in the derivative, we err on the parsimonious side. This is because a smaller number of clusters reduces the effect of noise and makes the recovered structure more meaningful. Because of the very high dimension, we show the results via a heat map, where each column corresponds to the values of one $g_{z,t}^1$ (or $\alpha^L_{t}$), and we delimit the clusters with dark vertical bars. If within a cluster the entries of all $g_{z,t}^1$ (or $\alpha^L_{t}$) are similar, then they are assigned similar colours on the heat map, and the cluster shows relatively uniformly across the $x$ axis.
\par
Figure \ref{fig:clusteringresults} suggests that we recover meaningful clustering structure, with generally well balanced clusters. This suggests that the TSOs choose the GSKs from a palette of values, and the variations within the clusters could be explained by expertise-based value adjustment, although a part of this could be due to noisy input data and the limitations of the numerical optimisation process, which does not guarantee global optimum. Nevertheless, the recovered clusters are meaningful, and the GSK choosing strategies discussed in \cite{Dierstein2017} support our findings.
\section{Conclusion}
Day Ahead Auctions in Power Markets that use the Flow Based Market Coupling methodology require linear constraints on all transmission lines to ensure that the auction leads to solutions that do not violate transmission load limits. In reality constraints corresponding to only a very small number of critical lines constraints are published. The missing data reduces market participants' risk management ability and ultimately leads to higher prices. 
\par 
We propose a methodology that recovers the electricity grid constraints via a mathematical optimisation problem using only publicly available data. The model reconstructs the grid structure and the underlying time dependent signals known as the Generation Shift Keys and Phase Angles, which are then transformed to the Power Transmission Distribution Factors and Remaining Available Margins, via mappings that are determined by the grid structure. Our results show good reconstruction quality and out of sample generalisation, and we recover meaningful clustering structure for the zonal GSKs and PAs, yielding insight into the operator's behaviour and suggesting that the GSKs could be chosen from a (fixed) palette of values up to small adjustments for daily characteristics. Our results have the potential to improve the risk management by day-ahead auction participants, leading to increased transparency,  reduced price inefficiency and higher social welfare.
\section*{Acknowledgment}
We would like to thank CFM London Power and Gas desk within Macquarie Group for partially funding this research.
\ifCLASSOPTIONcaptionsoff
  \newpage
\fi



%

%
\begin{IEEEbiographynophoto}{Ioan Alexandru Puiu} holds 
a B.Eng.\ degree from the University of Manchester and an M.Sc.\ in Applied Mathematics from the University of Oxford. He is currently pursuing a D.Phil. degree in Applied Mathematics at the University of Oxford. His research interests include machine learning, data analysis, optimisation and energy markets.
\end{IEEEbiographynophoto}
\begin{IEEEbiographynophoto}{Raphael Andreas Hauser}
is an Associate Professor at the Oxford Mathematical Institute, a Turing Fellow at the Alan Turing Institute in London, and Tanaka Fellow in Applied Mathematics at Pembroke College Oxford. He is a former recipient of the 
SIAM Activity Group on Optimization Prize. His research areas encompass numerical optimization, operations research, mathematical modelling and data science. 
\end{IEEEbiographynophoto}





\end{document}